# REBCO delamination by transverse electromagnetic stress due to screening current in magnetic field

Jun Lu, Jeremy Levitan, Yu Suetomi, Iain Dixon, and Jan Jaroszynski

*Abstract*— REBCO coated conductor has great potential to be used in ultra-high field magnets. Commercial REBCO tapes are strong in the longitudinal direction but prone to delamination by tensile stress in the thickness direction. For high field magnet applications, it is crucial to characterize delamination strength of REBCO conductor and better manage the transverse electromagnetic stress.

In this work, the electromagnetic stress in high magnetic fields by screen current is used to study the delamination behavior of commercial REBCO tapes. Screening currents are induced in REBCO by either ramping field or rotating sample in magnetic fields up to 35 T. The experimental results are presented. The prospect of using this method for quality assurance in large magnet projects is discussed.

## I. INTRODUCTION

RARE Earth barium copper oxide (REBCO) coated conductors is a high Tc superconductor that has shown good performance in ultrahigh magnetic fields [1]. As a result, it has found many promising applications in high field user magnets [2], nuclear fusion magnets [3], particle accelerator magnets [4], and high field NMR magnets [5]. REBCO is in tape form and mechanically strong in the longitudinal direction. However, its multilayer film structure is inherently weak under transverse tensile stress, and prone to delamination. This is a concern in ultrahigh field magnets where the accumulated transverse electromagnetic stress is high. Furthermore, in cases of epoxy impregnated magnets the transverse thermal stress due to differential thermal contraction can result in delamination which often leads to considerable critical current degradation [6]. Substantial efforts have been made to study the delamination strength of REBCO using various methods [7]-[12]. Recently, studies of REBCO delamination using electromagnetic stress in 19 T up to 2 kA current at 4.2 K have been reported [13], [14]. So far, there is a large variation in measured delamination strength values by different methods. To date the community has not yet found a consensus on the definition delamination strength that is most relevant to magnet design. Without a viable and reliable relevant testing method, the delamination strength cannot be defined, measured, monitored and eventually improved. For large magnet projects, quality assurance testing for the incoming REBCO tapes is vitally important. Establishment of a viable delamination strength test method and a delamination strength specification are necessary to screen out the tapes with low delamination strength.

In this work, we are developing a new test method using electromagnetic stress by screening current in magnetic field to determine the delamination strength. Many test methods use mechanical devices to apply stress at 77 K [7]-[12], which is different to the stress state in magnets operating at 4.2 K. In addition, these methods require careful sample preparation which might introduce measurement errors. Electromagnetic stress generated by screening current, on the other hand, is similar to the stress experienced in an operating magnet where the screening current persists and causes tape tilting [15]. It does not require a delicate sample preparation procedure and is not sensitive to the alignment of the measurement device, which make the measurement more reliable.

As shown in Fig. 1, a 4 mm wide sample is fixed on a metallic device with the REBCO layer facing outward and two ends of the samples were clamped down. In a magnetic field the induced screening current flows in a 'roof top' pattern as described by [16]. The electromagnetic force will push the upper trapezoidal-shaped region down onto the device and pull lower trapezoidal-shaped region up and possibly delaminated if the electromagnetic stress is greater than the delamination strength.

In this paper, we report our results in REBCO delamination strength measurement by transverse electromagnetic force generated by the induced screening current. We will compare our results with data available in the literature and discuss the correlation between these results and the 90° peel test results we performed previously [12]. The future development strategies will also be discussed.

This paragraph of the first footnote will contain the date on which you submitted your paper for review, which is populated by IEEE. This work was performed at the National High Magnetic Field Laboratory, which is supported by National Science Foundation Cooperative Agreement No. DMR-1644779, DMR-1839796, DMR- 2131790, and the State of Florida. *Corresponding author: Jun Lu*. Email:junlu@magnet.fsu.edu

Jun Lu, Jeremy Levitan, Yu Suetomi, Iain Dixon and Jan Jaroszynski are with the National High Magnetic Field Laboratory, Tallahassee, FL 32310 USA.

Color versions of one or more of the figures in this article are available online at http://ieeexplore.ieee.org

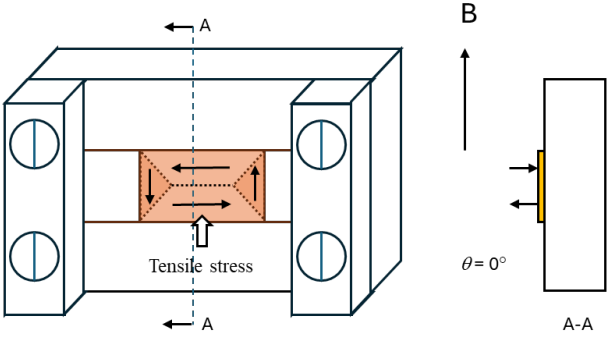

**Fig. 1.** A schematic of screening current and screening current generated delamination stress. The open arrow points to the lower trapezoidal region experiencing electromagnetic tensile stress.

## II. EXPERIMENT

*REBCO samples*

SuperPower (SP SCS4050) and Fujikura (FK FESC) tapes used in this work are 4 mm wide. Some properties of the samples are listed in Table I. The samples are typically 13 mm long. Some sample had their REBCO layer chemically removed except for a 5 x 3.5 mm² island.

TABLE I SAMPLES

| ID | Cu thickness (μm) | 77 K Ic (A) | 90° peel strength (N/cm) | RRR |
|---|---|---|---|---|
| SP-1093 | 20 | 137 | 0.12 | 46 |
| SP-1125 | 20 | 153 | 0.06 | 51 |
| SP-1155 | 20 | 178 | 0.82 | 63 |
| FK-164-01 | 5 | 189 | 0.71 | 21 |

*Screening current by ramping magnetic field*

The initial phase of the development is to simply use a probe with a stationary sample holder and the induction of screen current is realized by ramping magnetic field. The experiments were performed in a 35 T resistive magnet at the NHMFL (cell 8) at 4.2 K. After each field ramp to and from 35 T, the samples were taken out to visually check for delamination.

The magnitude of the stress in these experiments relies on the screening currents that flow at critical current density which is a function of magnetic field and field angle. Therefore, $I_c (B, \theta)$ at 4.2 K and up to 15 T is measured using torque magnetometry [20], [21]. The data is scaled up to 35 T and used in numerical calculation of critical currents and stress distributions.

The electromagnetic stress in transverse direction $\sigma_\perp$ can be calculated by,

$$\sigma_\perp = J_c B \cos\theta \qquad (1)$$

where $J_c$ is the current density in A/m, $\theta$ is the angle between the tape plane and the field. It is well established that $J_c \propto B^{-\alpha}$, where $\alpha$ is a power index which is 0.2 – 0.8 for $\theta = 0° – 90°$ at 4.2 K in SuperPower tapes [17]. So,

$$\sigma_\perp \propto B^{1-\alpha} \cos\theta \qquad (2)$$

Since $\sigma_\perp$ increases with $B$, ultrahigh magnetic field was used in this work to maximize achievable electromagnetic stress.

Our previous experiment showed that when $\theta = 0°$, the induced screening current by field ramping was very small [18]. This is because the component of the magnetic field perpendicular to tape surface is very small. Therefore, it is important to design a $\theta$ in this experiment that achieves relatively high stress. For this purpose, the induced screening current and the resultant electromagnetic stress were calculated numerically using COMSOL Multiphysics software. Fig. 2(a) plots the induced screening current density $J_c$ and electromagnetic stress for SP-1093 at 35 T as a function of $\theta$ after a 10 T/min field ramp. It is seen that the highest transverse stress is obtained at $\theta = 0.7°$ with the maximum electromagnetic stress of 9.2 MPa. Due to other engineering considerations, our experiment was conducted at $\theta = 3.5°$ as shown in the red data point where the stress is 7.1 MPa. Fig. 2(b) shows the induced screening current density along the sample width at $\theta = 3.5°$. The calculated screening current distribution is consistent with Bean's critical current model. Here an offset angle between REBCO ab-plane and the tape surface of 1.5° is measured by the rocking curve of x-ray diffraction [19] and considered in the calculations.

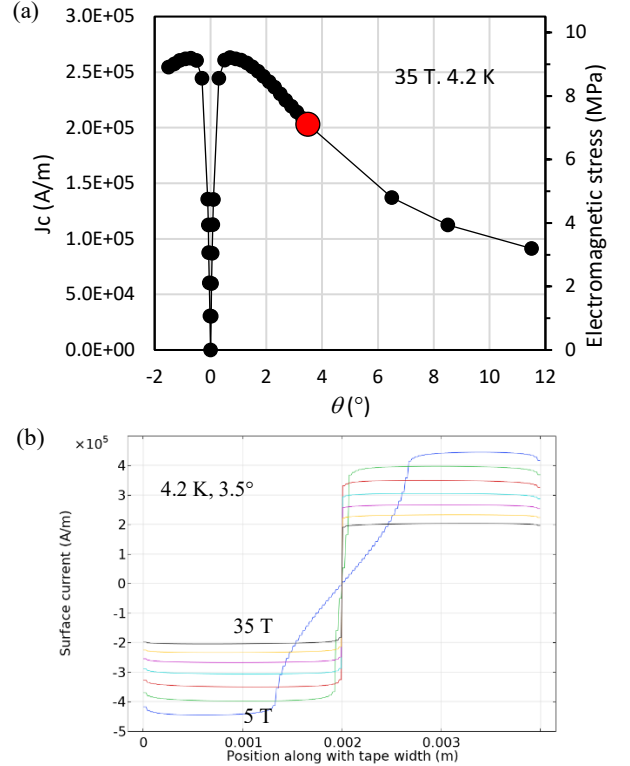

**Fig. 2.** (a) $J_c$ and electromagnetic stress vs. $\theta$ calculated for SP-1093. The field ramp rate is 10 T/min. (b) Screening current density distribution across width in 5 – 35 T, $\theta = 3.5°$.

*Screening current by rotating sample in magnetic field*

Despite the simplicity of the above measurement method, it lacks the ability to accurately determine the delamination strength. In the subsequent development a torque magnetometer probe [20] with a rotating sample stage was used in a 31 T resistive magnet at the NHMFL (cell 7). This enables the induction of screening current at $\theta = 0°$ and allows real-time measurement of the electromagnetic force by a load cell. A schematic of the probe is shown in Fig. 3.

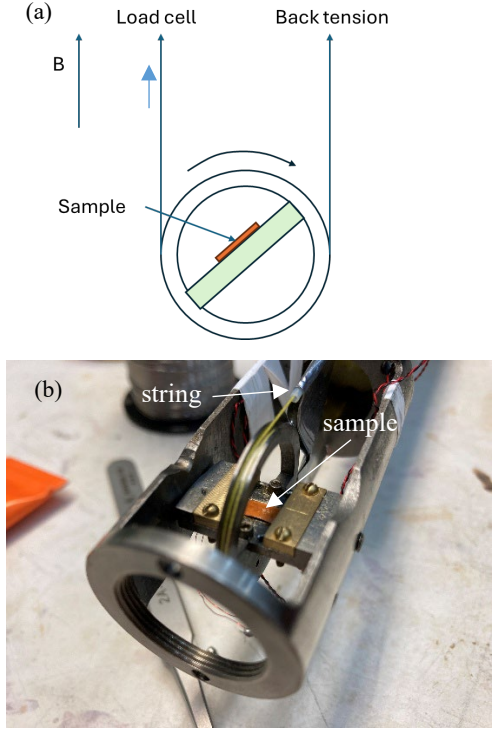

**Fig. 3.** (a) a schematic and (b) a picture depict the torque magnetometer. A string drives the sample stage rotation and connects to a load cell for force measurement.

III. EXPERIMENTAL RESULTS

*A. Ramping field at $\theta = 3.5°$*

The ramping field tests were performed on SP-1093 samples which were either clamped (Fig. 1) or soldered to the brass sample stage with the surrounding Cu intact. A 5 x 3.5 mm² island sample (without surrounding Cu) was also tested. None of the above samples were delaminated at the maximum field of 35 T. According to Fig. 2, this corresponds to electromagnetic stress of 7.1 MPa. It can be concluded that the delamination strength of SP-1193 is greater than 7.1 MPa.

*B. Tests using torque magnetometer probe*

In these experiments, the measured torque on sample $\tau(\theta)$ generated by electromagnetic force is converted to critical current density by [20],

$$J_c = 4\tau/(B\cos\theta \cdot l(1 - w/3l)) \quad (3)$$

where $l$ and $w$ are sample length and width respectively. Fig. 4 shows the measured induced critical current of SP-1125. The apparent oscillation at low $\theta$ is attributed to flux jumps due to the poor cooling of the sample where the sample is in 4.2 K helium gas and the fact that the sample was not in intimate contact with the sample stage. The maximum electromagnetic pressure at 30 T is 4.5 MPa as calculated by (1). No delamination was observed for the original SP-1125 or the one with Cu layer etched away and Ag layer exposed.

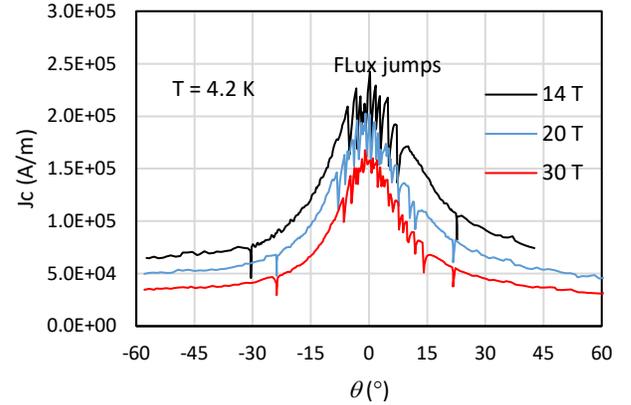

**Fig. 4.** Screening current vs. $\theta$ at 4.2 K of a SP-1125 sample that had the Cu removed (Ag layer on surface).

To observe delamination, SP-1155 was tested because it has significantly higher $J_c$ which can generate higher stress than SP-1125. This sample was etched to have a 5 x 3.5 mm² REBCO island on the substrate. In the process of forming the island, the edge Cu was removed. The result is shown in Fig. 5(a). A sudden drop in $J_c$ at about $\theta = 2°$ suggests that delamination occurred. The delamination was subsequently confirmed when the sample was taken out, only the substrate side of the sample with some residual REBCO remained on the sample stage. Fig. 5 (b) is a picture of the substrate side taken after the experiment. The color of the delaminated sample suggests that delamination occurred within the REBCO layer (black) as well as at the interface with the buffer layer (green). The delamination occurred at a $J_c$ corresponds to electromagnetic stress of 3.6 MPa as calculated by (1). For comparison, a Fujikura sample with a 5 x 3.5 mm² island was tested up to 4.95 MPa ($J_c = 1.65$ x $10^5$ A/m at 30 T) without delamination.

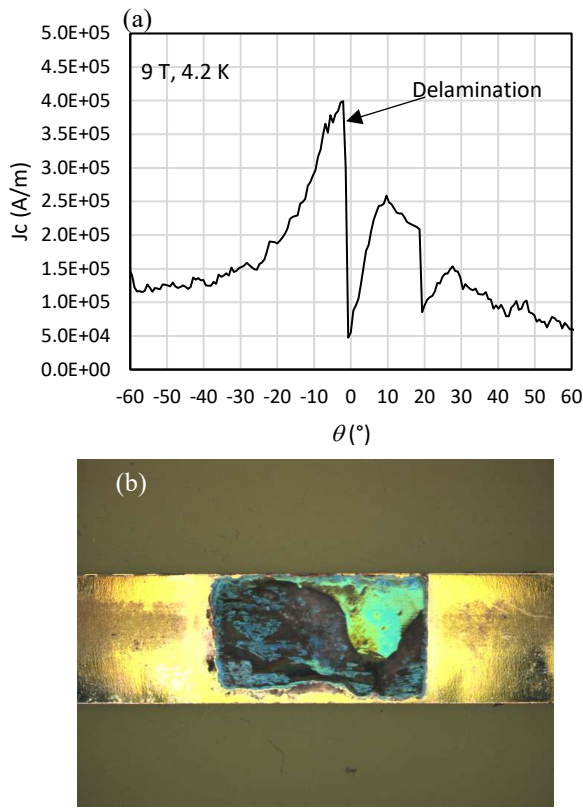

**Fig. 5.** (a) Critical current density $J_c$ of SP-1155 in a 9 T magnetic field. The sudden drop in $J_c$ indicates the occurrence of delamination. (b) the substrate side of the delaminated SP-1155.

## IV. DISCUSSIONS

It should be noted that there is only one region of the sample, i.e. the lower trapezoidal region in Fig. 1, experiences tensile transverse stress. We speculate that when delamination occurs, it starts in this region which leads to the delamination of the entire sample as shown in Fig. 5. As the delamination progresses, the screening current flow changes to unpredictable patterns very different from what is depicted in Fig. 1.

There are questions remaining. If screening current density is uniformly distributed, the electromagnetic stress accumulates through thickness and should reach maximum at the REBCO/substrate interface. If this is true, the delamination should occur near or at the REBCO/substrate interface. However, Fig. 5(b) show that the majority delaminated areas are in the REBCO layer.

Delamination strength of 3.6 MPa for SP-1155 is comparable to the values measured by electromagnetic stress of 3.9 – 5.5 MPa [13],[14]. But it is at least one order of magnitude lower than the tensile strength of YBCO bulk of over 100 MPa [22]. It is also several times lower than typical values measured by the transverse anvil test method [7]. On the other hand, it is much greater than the values obtained in cleavage test [8] where cleavage strength as low as 0.5 MPa was reported. Iin addition, the results in this work do not show correlation with 90° peel strength we tested previously and listed in Table I. Further investigations are needed to resolve these discrepancies and better understand REBCO delamination behavior, which includes the effects of the edge Cu and the sample temperature.

It is conceivable that delamination strength of REBCO varies between piece lengths even if they are made by the same process. Therefore, for a large-scale project it is prudent to perform quality assurance (QA) tests to screen out the tapes with low delamination strength. With further development, the method presented in this paper can be potentially used for QA testing. The limit of maximum stress may be overcome by stacking several pieces. The additional pieces of REBCO are used to provide additional electromagnetic stress that enables the testing of high strength samples without the need of high field resistive magnet. This, however, requires bonding the additional pieces, a process that could introduce additional stresses which might alter the strain state of the pristine sample. Further research and development on this are already underway.

## V. CONCLUSION

Electromagnetic stress introduced by screening current in high magnetic field is used to test delamination strength of commercial REBCO tapes at 4.2 K. Screening current is induced by either ramping field or rotating sample in magnetic field. The latter was achieved by using a torque magnetometry probe. Significant variations in delamination strength are observed. While one sample delaminated at 3.6 MPa, the other two samples did not delaminate at the applied stress of 5.0 MPa and 7.1 MPa respectively. This method, with further development, may be used in quality assurance testing to screen out REBCO tapes with low delamination strength.